\documentclass{PoS}
\usepackage[authoryear,square]{natbib}
\bibpunct{(}{)}{;}{a}{}{,}
 
\title{Strong gravitational lensing with the SKA}

\ShortTitle{Strong gravitational lensing}

\author{
\speaker{J. P. McKean}$^{1,2}$,
N. Jackson$^3$,
S. Vegetti$^4$, 
M. Rybak$^4$,
S. Serjeant$^5$,
L. V. E. Koopmans$^2$,
R. B. Metcalf$^6$,
C. D. Fassnacht$^7$,
P. J. Marshall$^8$,
M. Pandey-Pommier$^9$
\\
$^1$Netherlands Institute for Radio Astronomy (ASTRON), P.O. Box 2, 7990 AA Dwingeloo, The Netherlands; $^2$Kapteyn Astronomical Institute, University of Groningen, P.O. Box 800, 9700 AV Groningen, The Netherlands; $^3$Jodrell Bank Centre for Astrophysics, School of Physics and Astronomy, University of Manchester, Turing Building, Oxford Road, Manchester M13 9PL, United Kingdom; $^4$Max Planck Institute for Astrophysics, Karl-Schwarzschild-Strasse 1, D-85740 Garching, Germany; $^5$Department of Physical Sciences, The Open University, Milton Keynes, MK7 6AA, United Kingdom; $^6$Dipartimento di Fisica e Astronomia - Universita di Bologna, via Berti Pichat 6/2, 40127, Bologna, Italy; $^7$Department of Physics, University of California Davis, 1 Shields Avenue, Davis, CA 95616, USA; $^8$Kavli Institute for Particle Astrophysics and Cosmology, Stanford University, 452 Lomita Mall, Stanford, CA 94035, USA; $^9$CRAL-l'Observatoire de Lyon,  Universite de Lyon, 69561, France
\\
E-mail: \email{mckean@astron.nl}
}

\abstract{Strong gravitational lenses provide an important tool to measure masses in the distant Universe, thus testing models for galaxy formation and dark matter; to investigate structure at the Epoch of Reionization; and to measure the Hubble constant and possibly $w$ as a function of redshift. However, the limiting factor in all of these studies has been the currently small samples of known gravitational lenses ($\sim10^2$). The era of the SKA will transform our understanding of the Universe with gravitational lensing, particularly at radio wavelengths where the number of known gravitational lenses will increase to $\sim10^5$. Here we discuss the technical requirements, expected outcomes and main scientific goals of a survey for strong gravitational lensing with the SKA. We find that an all-sky (3$\pi$ sr) survey carried out with the SKA1-MID array at an angular resolution of 0.25--0.5 arcsec and to a depth of 3 $\mu$Jy~beam$^{-1}$ is required for studies of galaxy formation and cosmology with gravitational lensing. In addition, the capability to carryout VLBI with the SKA1 is required for tests of dark matter and  studies of supermassive black holes at high redshift to be made using gravitational lensing.}

\FullConference{
Advancing Astrophysics with the Square Kilometre Array\\
June 8-13, 2014\\
Giardini Naxos, Italy}

\begin{document}

\section{Introduction}

Gravitational lensing is the deflection of light from a distant background object ({\it the source}) by an intervening foreground mass distribution ({\it the lens}). If the surface mass density of the lens is sufficiently high, and if the source and lens are properly aligned, then multiple images of the background source are formed. The positions and distortions of these multiple images give important information about the mass distribution of distant lensing galaxies where velocity dispersions from spatially-resolved spectroscopy are either difficult or impossible. They also give a magnified view of the distant background source, probing a source population that is typically too faint to be detected with current instrument sensitivities. 

After the discovery of the first gravitational lens system B0957+561 \citep{walsh79}, in the Jodrell Bank 966-MHz quasar survey, the early history of strong lensing was dominated by VLA-based surveys, including the MIT surveys \citep{hewitt92} of mainly steep-spectrum sources and the JVAS/CLASS surveys \citep{myers03,browne03} of flat-spectrum radio sources. The majority of the currently known 36 radio-loud gravitational lens systems were found in this period. Further radio-based surveys became limited by the survey speed of high-resolution telescopes; because the typical separation of multiple images in gravitational lens systems, where the deflector is a massive galaxy, is $\sim1$~arcsec, gravitational  lens discovery programmes require large amounts of time on relatively oversubscribed interferometer arrays. Even with the recently upgraded JVLA and e-MERLIN, the survey speed is insufficient for large-scale searches. During the last 10 years, the advent of large-area optical imaging and spectroscopy surveys (e.g. SDSS) have found hundreds of gravitationally lensed star-forming galaxies \citep{bolton08} and quasars \citep{inada12}. The advent of wide-field surveys in the sub-mm with {\it Herschel} \citep{negrello10} and the SPT \citep{vieira13}, coupled with the resolution and sensitivity of ALMA, is also expected to increase the numbers of strong gravitational lenses by a few hundred. In total, there are around 300--500 gravitational lenses known from galaxy, galaxy-group and galaxy-cluster lensing, with about 10 per cent having a radio-loud (-bright) background source.

Gravitational lenses have been of extraordinary astrophysical value in many fields: for determining the radial mass density profiles of galaxies (e.g. \citealt{koopmans09}); the direct detection of sub-galactic scale substructure predicted by dark matter models for galaxy formation (e.g. \citealt{vegetti12});  measuring the mass of supermassive black holes in distant quiescent galaxies (e.g. \citealt{winn04}); measuring the stellar initial mass function in external galaxies by a combination of lensing and stellar population modelling (e.g. \citealt{auger10}); and by providing a one-step measurement of cosmological parameters, most notably the Hubble constant using time-delays (e.g. \citealt{suyu10}). Furthermore, the lens also magnifies the background source by factors of $\sim5$--100, which gives unique information about the structure of high redshift objects. This allows a source population that is intrinsically faint to be observed and studied, for example, the faintest known radio sources are currently gravitationally lensed objects (e.g. \citealt{jackson11}). Also, the magnification from the gravitational lens provides a high resolution (sub-kpc) view of galaxies at cosmologically distant epochs ($z \sim 1$--5; e.g. \citealt{deane13}). 

Most of these astrophysical applications have been limited by small-sample statistics, since only a minority of gravitational lenses are usually suitable for any particular study. However, we are on the verge of a huge increase in the number of gravitational lenses available for study in the radio with the SKA and in the optical/infrared with missions such as {\it Euclid} and the LSST. In this review, we show that a lensing survey with the SKA, coupled with the synergy provided by {\it Euclid} and the LSST, will open up a new parameter space, both in terms of statistics and in the discovery of rare and valuable gravitational lenses for studying galaxy formation, the high redshift Universe and cosmology. This review builds upon the Radio All-sky SKA Lensing (RASKAL) survey proposed by \cite{koopmans04} during the first SKA Science Workshop.

\section{Discovery space of SKA: The gravitational lensing statistics of continuum sources}

Surveys such as those envisaged with the SKA will be capable of increasing the sample of radio-loud gravitational lens systems from a few tens to tens of thousands relatively straightforwardly. In order to determine the actual numbers that are expected to be found, we need to evaluate the lensing probability, $\tau (z_s)$, of a given radio source,
\begin{equation}
\tau (z_s) = \int_0^{z_s} n(z)\sigma_x \frac{cdt}{dz}dz, 
\end{equation} 
where $z_s$ is the source redshift, $n(z)$ is the number density of the lenses, $\sigma_x$ is the cross-section of the gravitational lenses, and $cdt$ is the path length to the background source (e.g. \citealt{turner84}). By integrating the lensing probability over all of the sources in the parent sample, we can determine the total number of sources that are gravitationally lensed. If we know the number counts of the parent sample as a function of flux-density and redshift, then we can calculate the number of gravitational lenses expected within a flux-limited survey. To zeroth order, gravitational lensing is a rare event; a high-redshift object has a $\sim10^{-2}$--$10^{-3}$ chance of lying within the Einstein radius of an intervening galaxy-scale mass distribution (depending on the source population redshift and magnification bias). For cluster lensing the probability is even smaller, at about $\sim10^{-4}$. Therefore, surveys of a few hundred million high redshift objects should in principle give access to a few hundred thousand lenses, although recognition difficulties will mean that the actual number discovered will be a factor 10--100 less than this (see Section 3).

In order to estimate the lensing potential of the SKA surveys, we first require a model that describes the lens and source populations. To calculate the number density of lenses, we have used the light-cones generated by \citet{kitzbichler07} coupled with semi-analytic prescriptions \citep{delucia07} from the Millennium simulation \citep{springel05}. We have assumed that each potential lens has a velocity dispersion that is related to its luminosity via the Faber-Jackson relation, with a normalisation that gives for an $L_*$ galaxy a velocity dispersion of $\sigma_* = 190$~km$\,$s$^{-1}$. We have then scaled the $L_*$ luminosity using the lensing statistics from the CLASS survey \citep{myers03}; this is done by requiring that an initial sample of radio sources with flux densities $>1$~mJy, and  differential source counts with a slope of $-2$ \citep{mckean07} produces 22 gravitational lenses, as observed by CLASS \citep{browne03}. This has been done by assuming the CLASS selection criteria, namely that selected gravitational lenses should have flux ratios $<10$:1 and an image separation $>300$~mas. To first order, this calibration removes the effects of cosmic variance. The SKADS database \citep{wilman08} was used to provide the population of background radio sources. This population is a best-guess extrapolation of known radio surveys down to the $\mu$Jy level (e.g. \citealt{condon12}), and will almost certainly be superseded once the SKA surveys begin in earnest. The exact nature of the very faint ($<0.1~\mu$Jy) population is important, because the magnification provided by lensing (factors 5--100) yields many more gravitational lens systems in populations with a steep source count, a phenomenon used to advantage in sub-mm surveys (e.g. \citealt{negrello10}).

\begin{figure}
\includegraphics[width=\textwidth]{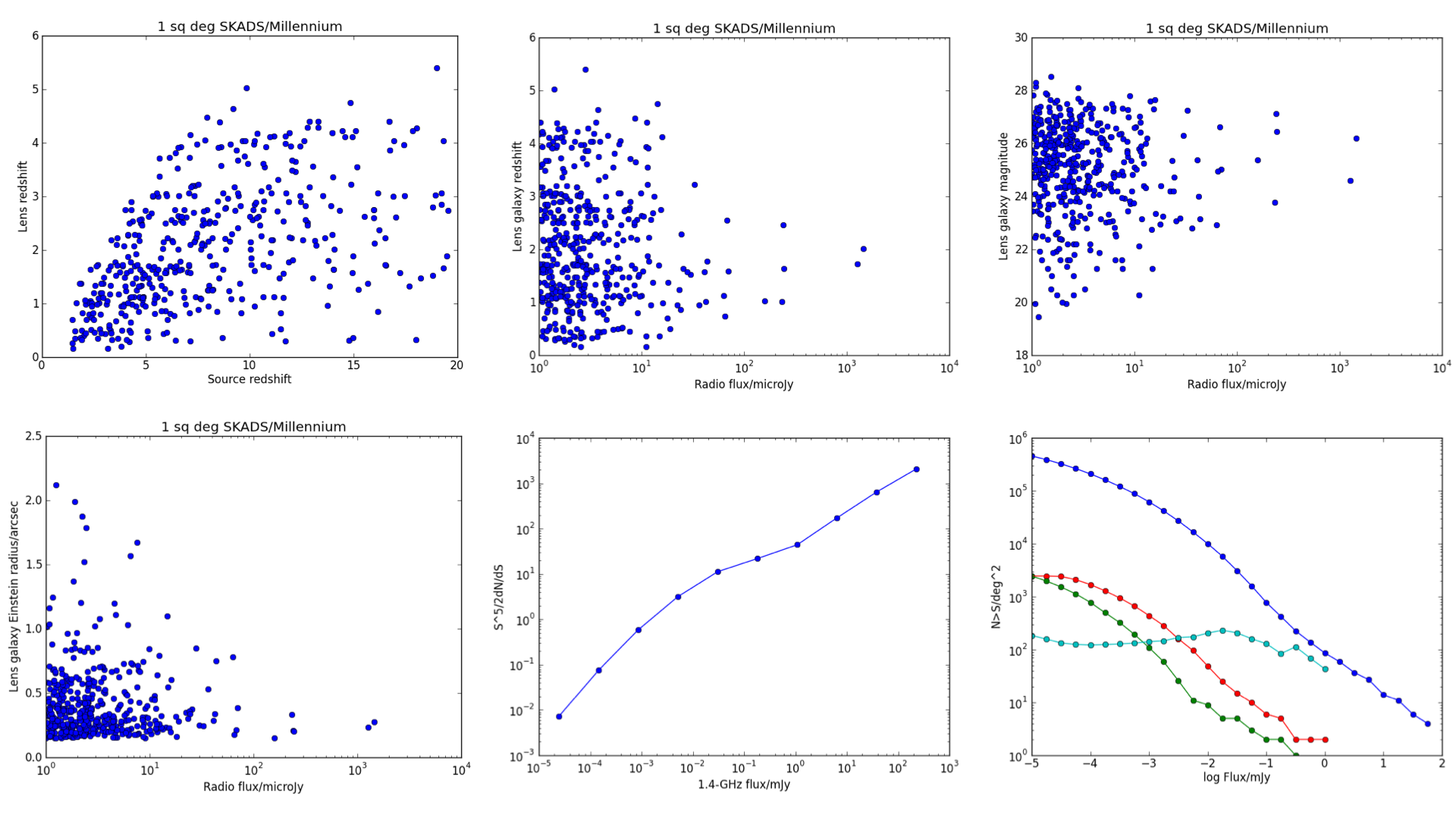}
\caption{The results from the SKA gravitational lensing simulations. {\it Upper left}: Lens galaxy redshift versus source galaxy redshift. {\it Upper middle}: Lens galaxy redshift versus source galaxy flux density. {\it Upper right}: Lens galaxy magnitude versus source galaxy flux density. {\it Lower left}: Lens galaxy Einstein radius  versus source galaxy flux density. {\it Lower middle}: The predicted differential number counts normalised to Euclidian from the SKADS simulation. {\it Lower right}: The total integrated number counts as a function of flux density for all sources (dark blue), the lensed sources (red), the lensed sources using the unlensed flux-density (green) and the lensing incidence (cyan).}
\label{sim}
\end{figure}

In total, $4.56 \times 10^5$ radio sources from a 1 square degree area of SKADS and with a 1.4-GHz flux density $>1~\mu$Jy were extracted and randomly placed in the Millennium area. Any galaxy within 1~arcsec was considered to be a potential lens; the lensing status was then evaluated for each such pair. The results from the simulation are shown in Figure \ref{sim}. In the 1 square degree of SKADS used, 440 radio sources with a total radio flux density $>1~\mu$Jy were found to be gravitationally lensed and have image separations $>0.3$~arcsec. Of these lensed sources, 49 have total flux-densities $>10~\mu$Jy, and 10 have total flux-densities $>50~\mu$Jy. These calculations clearly show that for a fixed observing time it is most efficient to carry out a shallow, wide-field survey for gravitational lenses with the SKA. As we will demonstrate in Section 3, a survey with an rms of $\sim 3~\mu$Jy~beam$^{-1}$ (2 year duration), and a detection threshold of about 15$\sigma$ for lensing, yields about $\sim$5 lenses per pointing (0.5 square degrees) using the full 250 dish SKA1-MID array. These systems will contain lensed objects typical of the radio population at a slightly fainter flux density level of a few tens of $\mu$Jy, in other words, approximately evenly distributed between lensed AGN and lensed starbursts. 

The simulations also show the gravitational lenses that may be available from surveys with the SKA2, which will be up to a factor of 5--10 deeper than with SKA1-MID. These are potentially far more exotic objects, with the lensing galaxies at redshifts 1--4, extremely faint, very high redshift background sources, and in very large numbers despite the unfavourably flat faint radio source count of SKADS; the lensing rate approaches $\sim1$:150 due to the high redshifts (median $z \sim 3$) of the background source population. The details of these samples should be taken with caution because they rely on uncertain extrapolations. In particular, the likelihood of a large population of radio sources at $z>10$ is low. What these simulations do illustrate, however, is the unique power of a gravitational lensing survey to probe any faint, high redshift radio source populations that may exist, if we know the detailed properties of the lensing galaxies. The latter is currently unknown, but is an area of very active study that will progress greatly over the next 5--10 years. For example, a deep (sub-$\mu$Jy) small area continuum survey with the SKA1-MID array would directly detect this population of magnified and exotic high redshift objects, accessing the type of objects that will be seen routinely with the SKA2. Alternatively, if the predicted number of lenses is not found, then this would mean that the faint radio source population has a much different luminosity function than is currently expected. However, this assumes that the challenges related to the lens identification and confirmation process are well understood, which is typically the case for lens surveys are radio wavelengths where the lens selection function (resolution, sensitivity, flux-ratios; e.g. \citealt{myers03}) is well defined.

The major uncertainties in the number of gravitational lenses that will be potentially available for study in any SKA survey are the extrapolation of the radio source counts to very faint flux density levels and high redshifts, and 
the detection efficiency. The former uncertainty is negligible at the level of a few tens of 
$\mu$Jy (which is most relevant for an SKA1 survey), but is very probably a factor of a few at the 1-$\mu$Jy level, and likely an order of magnitude for the $z>5$ radio source population. This uncertainty dominates the estimates for any SKA2 survey. The detection efficiency has an uncertainty that is likely to be of a factor of a few, although more detailed blind simulations are needed to reduce this uncertainty.

\section{Gravitational lens candidate identification and confirmation techniques}

The main observational challenge will be the recognition of genuine gravitational lenses among the large numbers of false positives corresponding to radio sources with intrinsic structure; the key criterion is angular resolution, although frequency coverage and image sensitivity also play a role. Known galaxy-scale gravitational lenses have image separations between $\sim0.3$--4.5 arcsec, with the typical image separation being $\sim1$~arcsec (e.g. \citealt{mckean05}). From our SKA lensing simulations, the predicted image separation distribution for sources with a total flux-density $>50~\mu$Jy has a range between 0.3 and 1.4 arcsec. Therefore, an angular resolution of between 0.25--0.5 arcsec is required for identifying most of the lens candidates for follow-up imaging. It is for this reason that gravitational lensing surveys will only be possible with the SKA1-MID array.

\begin{figure}
\includegraphics[width=\textwidth]{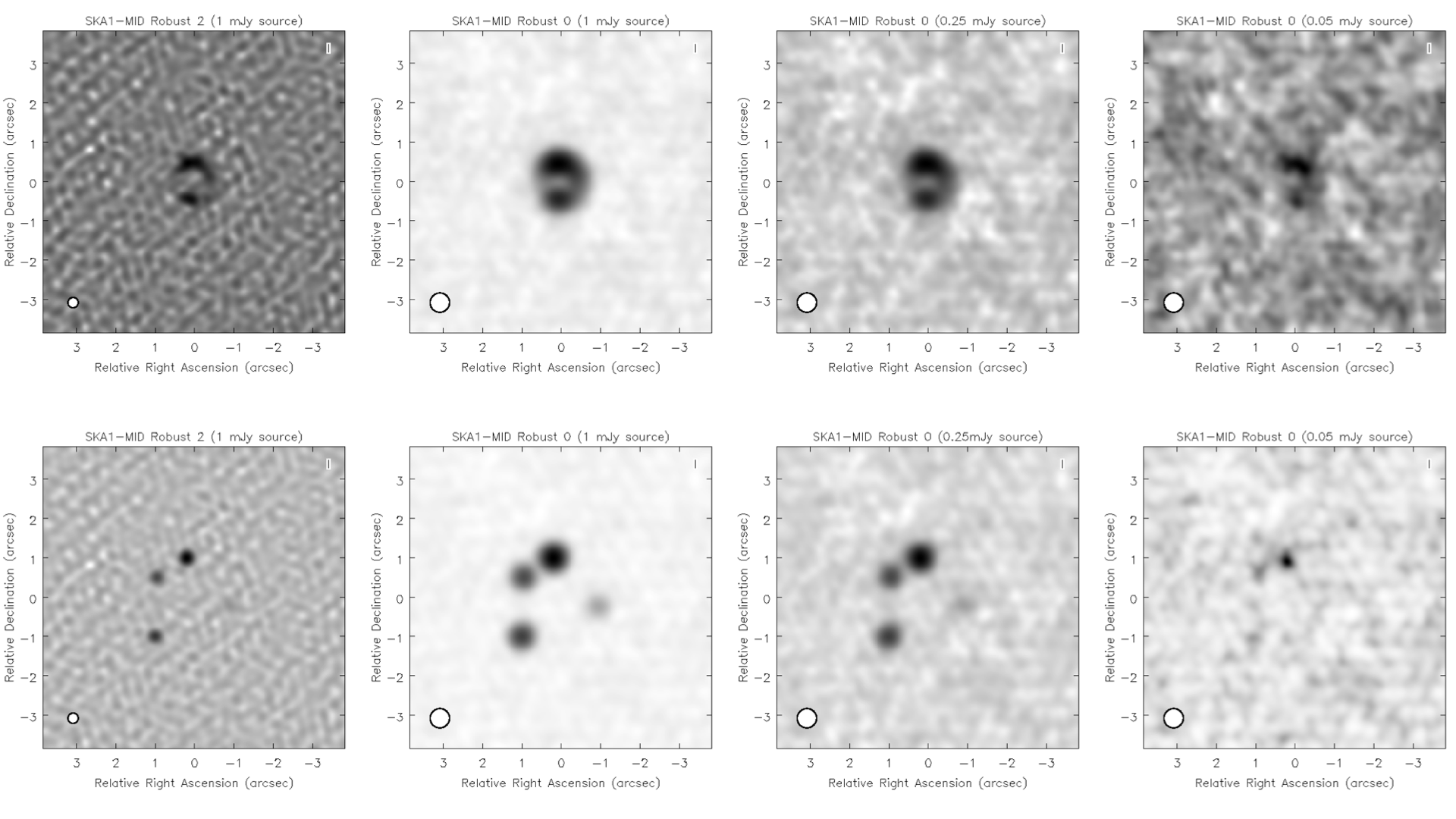}
\caption{Simulations of an SKA1-MID snapshot observation of an extended radio galaxy (top) and an AGN (bottom) with total flux-densities of 1, 0.25 and 0.05 mJy (left middle, right middle, right) using Briggs weighing Robust $= 0$ (0.5 arcsec beam size) and for a total flux-density of 1 mJy (left) with Robust $=2$ (0.25 arcsec beam size).}
\label{imaging-sim}
\end{figure}

To simulate the expected imaging quality for a gravitational lens search, we have taken the array configuration for SKA1-MID and generated mock gravitational lensing data based on an extended star-forming galaxy and a compact AGN; for both types of object, we have used {\it total} source flux-densities of 1, 0.25 and 0.05 mJy. From Figure \ref{imaging-sim}, we see that sensitive observations (Briggs weighting; Robust = 0) with SKA1-MID have sufficient sensitivity ($\sim3~\mu$Jy~beam$^{-1}$) and angular resolution ($\sim0.5$~arcsec) to detect galaxy-scale gravitational lenses with extended sources. We find that it is possible to achieve an angular resolution of 0.25 arcsec (for Robust = $2$), but as expected, this increases the image rms by a factor of $\sim5$, leading to the fainter extended radio sources having a surface brightness below the detection limit. Nevertheless, observations at Band 2 (0.95--1.76 GHz) with SKA1-MID can be used to select gravitational lensing candidates. An all-sky (3$\pi$ sr) survey with SKA1-MID, reaching a sensitivity of $3~\mu$Jy~beam$^{-1}$ could potentially find $\sim3\times 10^5$ gravitational lens systems with image separations $> 0.3$~arcsec.

\begin{figure}
\centering
\includegraphics[width=5.9cm]{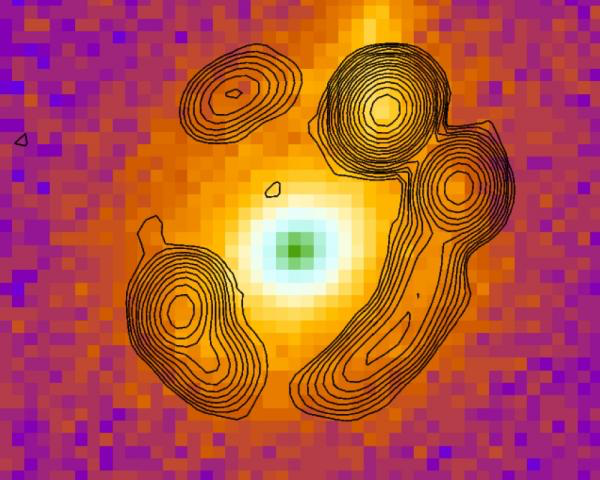}
\caption{High resolution {\it HST} optical imaging of the gravitational lens CLASS~B0631+519 with the MERLIN 1.7 GHz radio contours shown overlaid. The optical data only detect the two lensing galaxies, whereas the radio data are only sensitive to the emission from the background source. Together, these two datasets clearly show that CLASS~B0631+519 is a gravitational lens \citep{york05}.}
\label{imaging-radio-hst}
\end{figure}

The resulting recognition problem (candidate to confirmed lens) will be considerable, but can be eased by adding additional information. The first is the large bandwidth to be used in the SKA1-MID survey, which will mean that some spectral index information will be available to discriminate between multiple images of the same object and, for example, a non-lensed source consisting of a flat-spectrum core and steep-spectrum jet or lobe emission. In addition, follow-up observations of candidates at Band 5 (4.6--13.8~GHz) will give a more robust measurement of the spectral index of the components and provide higher angular resolution imaging of the candidates (0.06--0.20 arcsec; Robust $= 0$ weighting) to determine if the surface brightness of the images is conserved, as is required by gravitational lensing. The second piece of information will come from surveys at other wavelengths, in particular those of {\it Euclid} and the LSST. Cross-correlating the SKA gravitational lens candidates with the high resolution optical data will determine whether there is a massive lensing galaxy at the expected position (see Figure \ref{imaging-radio-hst}) and provide a photometric redshift estimate of the gravitational lens. Even if neither the SKA1-MID and {\it Euclid}/LSST surveys are capable of detecting gravitational lenses on their own, experience from CLASS shows that the use of two surveys at different wavelengths (radio and optical) gives a very much greater discrimination between multiply-imaged sources and those that display intrinsic structure. 

Using the predicted number counts and image separation distribution from our SKADS simulation, we have estimated the number of gravitational lenses that should be straightforwardly identified from the SKA1-MID survey presented above with minimum follow-up. Here, we only consider those lensed images with separations $>0.5$~arcsec and total flux-densities $> 1$~mJy. We find a conservative estimate of about $\sim10^{4}$ new radio-loud gravitational lens systems would be found ($\sim1$ per 3 square degrees), which is a factor of over $10^2$ more than is currently known. We note that an initial pilot survey, that covers the same sky area as the {\it Euclid} deep fields (15000 deg$^2$) and to an rms of 6 $\mu$Jy~beam$^{-1}$, would conservatively identify around 2000 strong gravitational lenses in 3 months observing time. An SKA1-MID array that is only 50 per cent complete would still be useful for finding gravitational lenses, providing the sensitivity and length of the baselines remains unchanged and a wide area survey is carried out. The situation will improve dramatically with the SKA2, where the improved sensitivity, particularly on the long baselines will allow the potentially large population of $\sim10^5$ faint lensed radio sources to be straightforwardly identified in the survey imaging data.

\section{Searches for gravitationally lensed spectral line sources} 

An alternative method to efficiently find new gravitational lens systems with the SKA is to select candidates via magnification bias; here a specific property of the source results in very steep number counts. For example, \cite{negrello10} successfully used this technique to select a sample of strong gravitational lenses by using a $500\,\mu$m flux density limit and exploiting the steep slope of sub-mm number counts to obtain a $\sim100$~per cent reliable strong gravitational lens sample. \cite{gn12} extended this technique to the high-luminosity end of the sub-mm luminosity function. The HI mass function is expected to broadly follow its local form of a Schechter function (e.g. \citealt{abdalla10}). Therefore, a wide-field SKA HI survey, as proposed by \cite{abdalla10}, should also be amenable to magnification bias at the high-mass end of the HI mass function.

For the detailed formalism used for calculating differential magnification probability distributions, $p(\mu,z){\rm d}\mu$, for a strong gravitational lens with magnification $\mu>2$, see \cite{perrotta02,perrotta03}. It is expected that the high magnification tail has the form $p(\mu,z)=a(z)\mu^{-3}$ for some function $a(z)$, regardless of the lens population. The $a(z)$ function depends on the nature and evolving number density of the gravitational lenses. \cite{blain96} and \cite{perrotta02,perrotta03} consider several options for galaxy lens populations. In the latter, the mass spectrum follows the \cite{sheth99} formalism.  As for the continuum case (Section 2), we assume a singular isothermal sphere density profile for the lens population and assume a non-evolving model, but normalised to the Perotta et al.\ predictions at an arbitrary redshift of $z=1$. 

A more significant uncertainty in the lensing predictions is the maximum magnification caused by the finite source sizes. We follow \cite{perrotta02} in spanning the range of plausible maximum magnifications with $\mu\leq10$ and $\mu\leq30$, implying source characteristic radii of $\sim1-10~h^{-1}$\,kpc; note that if the assumed source sizes are larger than this then the maximum magnifications will be lower, but it is expected that galaxies will be systematically more compact at higher redshifts (e.g. \citealt{gun1972}) so our assumed sizes are likely fair.  Several higher magnification events are known (e.g. \citealt{swinbank10}), but we conservatively neglect this more extreme population. Note that the surface density of strong gravitational lenses increases with maximum magnification. The choices of maximum magnification here is more conservative than that of, for example, \cite{blain96} who used $\mu\leq40$, and \cite{lima10} who used $\mu\leq100$.

\begin{figure}
\centering
\includegraphics[width=\textwidth]{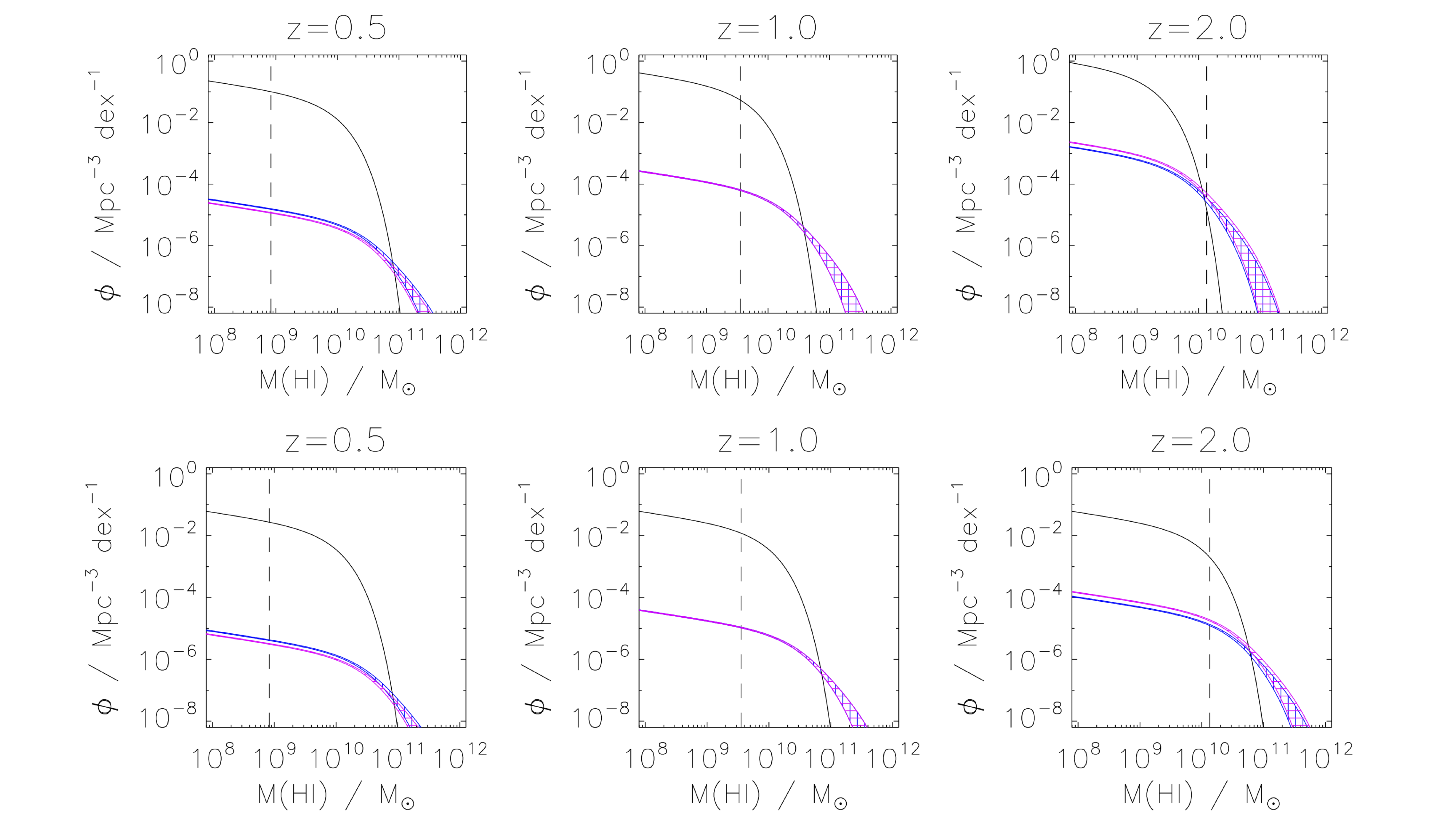}
\caption{The HI mass function (black line) at redshifts
  $z=0.5,1,2$, assuming no evolution from the $z=0$ mass function
  (top) and assuming an evolving mass function (Bottom). The no-evolution lens population is shown in blue, and
  the singular isothermal sphere lenses of \cite{perrotta02,perrotta03} are in pink.  The hatched regions show the
  range of maximum magnification $\mu\leq10$ and $\mu\leq30$.  Note
  that at $>12~M_*$ the observed population is dominated almost
  entirely by gravitationally lensed HI galaxies.}
  \label{hi-sim}
\end{figure}

Figure~\ref{hi-sim} shows the results of applying this strong gravitational lensing formalism to a non-evolving HI mass function \citep{zwaan03}, and to the evolving mass function of \cite{abdalla10}.  Selecting gravitational lens candidates as those with HI masses of $>12~M_*$ in an HI spectral line survey, should yield a sample of between 0.5--5 gravitational lenses per square degree, but with a $\sim100$ per cent selection efficiency and the key advantage that the source redshifts are known. Further details, and additional calculations for the Dark Energy Grism survey with {\it Euclid}, are presented by \citet{serjeant14}. 

\section{Science capabilities with SKA}

In this Section, we review the main astrophysical results that will come from the strong gravitational lensing programme of the SKA. These will be a combination of results that will exploit the large statistical sample of gravitational lenses, or will use smaller samples of rare gravitational lenses for investigating galaxy formation and cosmology.

\subsection{Galaxy formation studies} 

The hierarchical model for galaxy formation predicts that structure in the Universe formed through the mergers of smaller mass systems, resulting in the tapestry galaxies, clusters of galaxies and large-scale structure that we observe today. A crucial result of this model is that the stellar and gas components of these structures should be embedded within an extended dark matter halo (e.g. \citealt{navarro96}). The form of the radial mass-density profile of the resulting dark matter haloes and the level of low mass substructure within them are both sensitive to the galaxy formation model and the energy of the dark matter particle.

Over the last decade, gravitational lenses have been used to place important constraints on the mass distributions for galaxies at cosmological distances and have tested these predictions from hierarchical galaxy formation models. At radio wavelengths, high resolution imaging of gravitationally lensed extended radio sources at mas-scales with VLBI have constrained the inner (within 5--20~kpc) radial mass density profiles for a handful of lens galaxies (e.g. \citealt{cohn01,wucknitz04}), finding that they are close to isothermal (i.e. $\gamma = 2$ where $\rho (r) \propto r^{-\gamma}$). However, the most significant results have come at optical wavelengths, where high resolution imaging with the {\it HST} of the Einstein rings and arcs of extended blue star-forming galaxies, coupled with stellar kinematics (velocity dispersions and velocity fields) for the lens, have determined radial mass density profiles for a homogenous sample of 58 early-type galaxies at $z\sim0.2$ \citep{koopmans09}. Again, these results find that massive galaxies have on average isothermal density profiles ($\gamma= 2.09\pm 0.02$). 

An important constraint on galaxy formation models from gravitational lensing comes through the detection of low mass substructure around more massive galaxy-scale haloes. The cold dark matter model for galaxy formation predicts a substructure mass function, $dn \propto dm\,m^{-1.9}$ (e.g. \citealt{diemand07}), and a mass fraction in substructure of $<1$\%. However, the lack of observed satellites found around our own Milky Way suggests that the mass function may be much flatter than thought, which could be due to our Milky Way being a unique case, the substructures being there in the expected abundance, but are dark, or due to a differing physics of the dark matter (e.g. warm dark matter; self-interacting dark matter). Gravitational lenses can be used to test galaxy formation models on these small-scales because the surface brightness distribution of the observed images depends on the form of the potential and they can be (very) sensitive to local perturbations in the mass model due to small clumps of dark matter in the parent halo (or along the line-of-sight). There are two main applications of this technique. The first is {\it gravitational imaging} \citep{koopmans05,vegetti09}, which uses the information contained within extended arcs to detect the presence of individual substructures and measure their mass properties. This technique has been used to detect substructures with masses as low as $2\times10^{8}~M_{\odot}$ out to redshift $z= 0.881$ from infrared adaptive optics imaging at $\sim65$~mas resolution \citep{vegetti12}. The second method uses {\it flux-ratio anomalies} (e.g. \citealt{metcalf01,dalal02}), which occur when the flux-density of one or more point images are different from what is predicted by a smooth mass model. This method is sensitive to the total amount of substructure at the Einstein radius of the lens, and can be used to statistically detect a substructure population down to $\sim10^4~M_{\odot}$. Here, observations at radio wavelengths are important because the flux-ratios are less effected by microlensing from stars in the lensing galaxy \citep{koopmans03} and extinction due to dust is not an issue.

The SKA, when combined with optical telescopes like {\it Euclid} and the LSST, will revolutionise studies of the mass distributions of galaxies by increasing the overall sample sizes and will allow measurements to be made for samples over a wide range of galaxy mass, type, redshift and environment. The main results that will come from the SKA strong gravitational lensing sample are as follows.

\begin{figure}
\includegraphics[width=\textwidth]{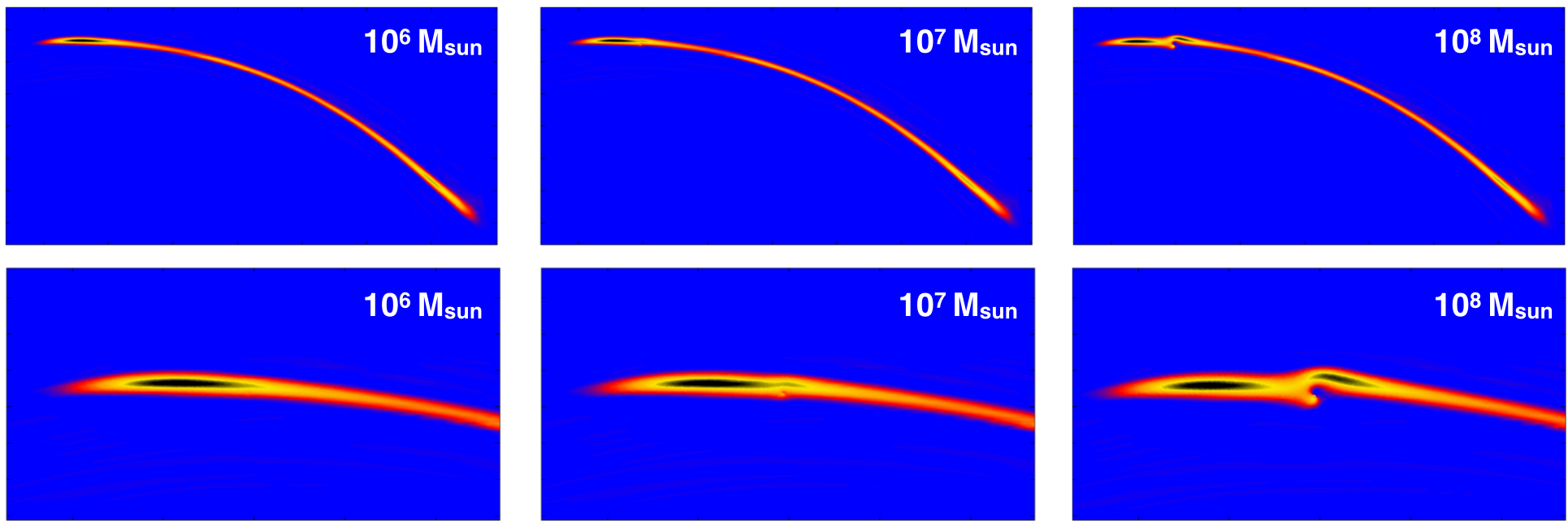}
\caption{Simulations of the effect low mass dark matter haloes have on arc surface brightness distributions for a $10^6$~M$_{\odot}$ (left), $10^7$~M$_{\odot}$ (middle) and $10^8$~M$_{\odot}$ (right) substructure. The top row of images show the full gravitational arc, detected on VLBI scales with an SKA1-MID+VLBI array, and the bottom row show a zoomed in view of the region most affected by the substructure. The $10^7$~M$_{\odot}$ and $10^8$~M$_{\odot}$ substructures can be seen directly in the data, whereas the $10^6$~M$_{\odot}$ substructure would need to be detected using the gravitational imaging technique.}
  \label{vlbi}
\end{figure}

\begin{figure}
\includegraphics[width=\textwidth]{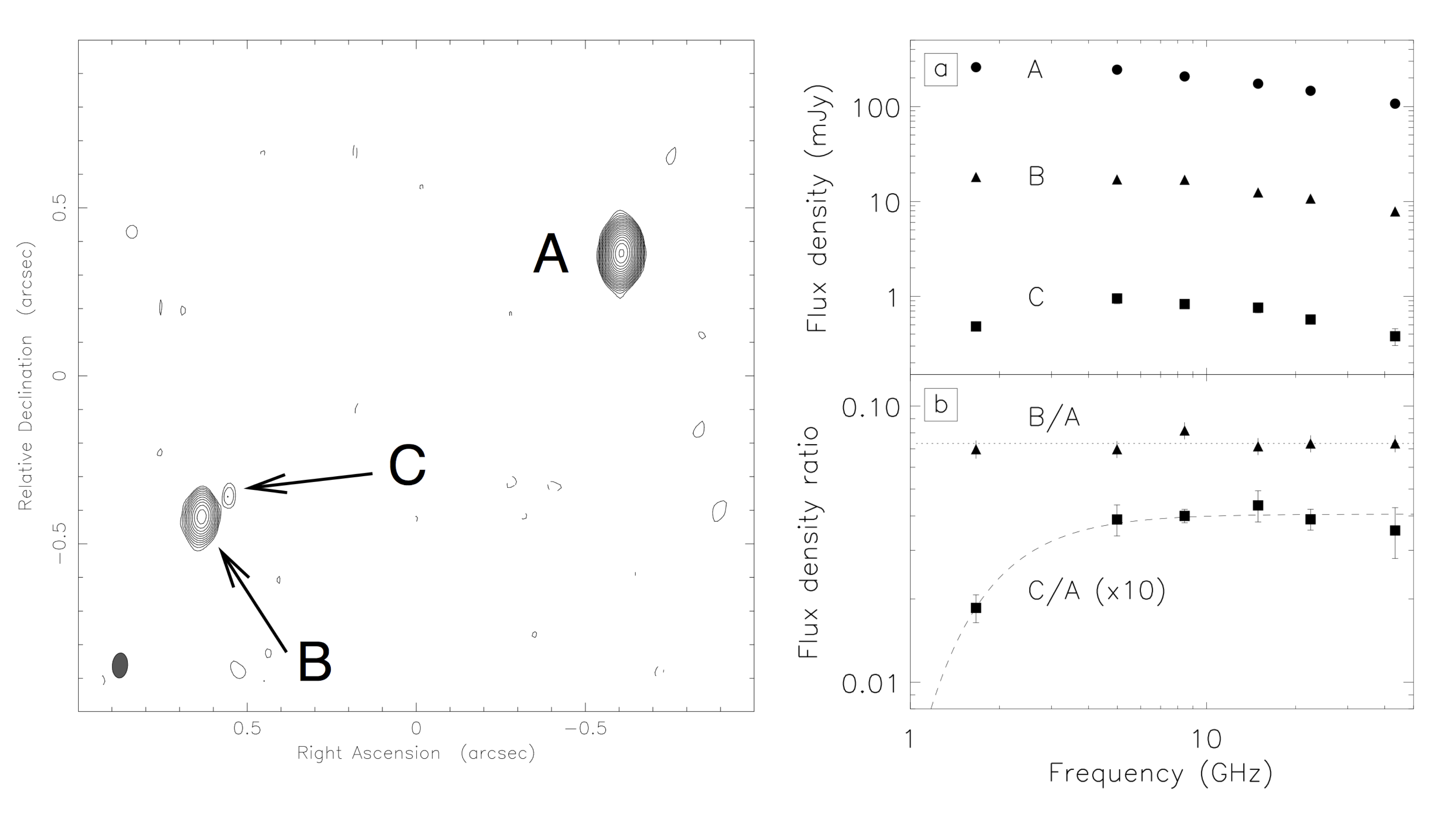}
\caption{The gravitational lens system J1632-0033 is the first and only gravitational lens system to have a central lensed images for a galaxy-scale system. Images A and B are of the redshift 3.42 lensed radio sources, whereas image C is the core lensed image. The spectra of the three images are identical are high frequencies, since propagation effects (free-free absorption within the lensing galaxy) are less severe \citep{winn04}.}
\label{central}
\end{figure}

\begin{itemize}
\item The main constraints on the radial mass-density profiles of structure will come from high resolution imaging taken in Band 5 (4.6--13.8~GHz) with the SKA1-MID array, which can potentially be coupled with HI velocity field measurements of lenses at low redshift ($z<0.4$) using spectral line observations in Band 2; combining the constraints from the lensing and dynamics breaks the degeneracies between mass models that are inherent in both methods individually (e.g. \citealt{barnabe09}). The completion of the SKA2 will allow a much larger population of lenses at higher redshift to be studied in this way, which will determine whether there is any evolution in the mass properties of galaxies with epoch and provide a key observational constraint to galaxy formation simulations.

\item The very precise measurement of the projected mass within the Einstein radii of galaxies ($\sim10$~kpc) from strong lensing can be used, with optical and infrared measurements, to independently determine the stellar mass and place constraints on the stellar initial mass function (e.g. \citealt{auger10}). This requires both the redshift of the lens and source to be known, and the modelling of the SKA1-MID array data to determine the Einstein radii of the lenses. The SKA strong gravitational lenses will provide a large statistical sample to investigate variations of the stellar IMF with galaxy type and mass when coupled with the optical/IR imaging provided by {\it Euclid} and the LSST.

\item The SKA1-MID array has the potential to provide important constraints at the low end of the dark matter halo mass function using strong gravitational lensing, which would independently test models for the dark matter particle. The main requirement for the gravitational imaging method is the angular resolution of the data (source structure and signal-to-noise ratio are also important). Follow-up observations with SKA1-MID in Band 5 of those lenses with extended radio sources will give an angular resolution of 0.03--0.1~arcsec, similar to the resolution provided by the {\it HST} and ground based adaptive optics telecopes, and so should give detections of substructures down to $\sim10^{8}~M_{\odot}$ for a large sample of lens galaxies. However, the most important constraints to the dark matter halo mass function will come from observations on VLBI-scales since the mas angular resolution will be sensitive to structures at the $\sim10^{6}~M_{\odot}$ level (see Figure \ref{vlbi}), a mass regime where various models for the dark matter particle strongly differ. Therefore, an SKA1-MID array that is capable of VLBI observations with, for example, the EVN, is required for this unique science goal of the strong gravitational lensing programme to be achieved.

\item It will also be possible to place constraints on the substructure mass function with the SKA1-MID array using those gravitational lenses with compact (point) emission from AGN (see Figure \ref{imaging-sim}). It is expected that there will be $\sim 300$ four image gravitational lens systems from the SKA1-MID survey that can be used for this; a factor of about 50 more than are currently known. This will require the precise measurement of the flux-ratios of the point-like images; this will be done through monitoring the systems with the SKA1-MID Band 5 to achieve the desired angular resolution and to determine if the fluxes have been changed by intrinsic variability and the gravitational lensing time-delay (the latter can be used to determine the cosmological parameters, see below).

\item The sensitivity of the SKA will allow the routine detection of highly de-magnified core lensed images that are predicted to form at $<100$~pc from the centre of the lensing galaxy (for shallower than isothermal lensing potentials). Such images probe the central mass distribution and can be used to constrain the central density profile where dark matter is insignificant, but the influence of both the central supermassive black hole (SMBH) and the central stellar cusp are important (e.g. \citealt{mao01}). Understanding the central regions of galaxies is important because they are directly relevant to the question of how the SMBH influences (and is influenced by) the process of galaxy formation. This central question has been debated since the discovery of the relation between the SMBH mass and the bulge stellar velocity dispersion on much larger scales \citep{Ferrarese00}. However, the detection of core-lensed images is extremely rare due to their very high de-mganification ($\sim10^{-4}$--$10^{-5}$) relative to the other lensed images. There is currently only one confirmed case \citep{winn04}.

Targeted observations with SKA1-MID should allow the detection of core-lensed images in about 100 systems, compared to the 20--30 accessible to e-MERLIN, provided that an angular resolution of about $<100$ mas can be achieved, as is expected from observations at Band 5 with the SKA1-MID. The large frequency coverage of Band 5 is also required to confirm that the emission is from the central image and not the lensing galaxy (see Figure \ref{central}). With the SKA2, measurements of the central-image properties of thousands of lens systems will be possible, and hence will provide a census of central potentials in many different lensing galaxies as a function of redshift, mass and environment. With these sensitivities, non-detections become just as interesting as detections because they typically indicate SMBH masses greater than the $M$--$\sigma$ relation, unusually steep central density profiles, or both. Moreover, in combination with VLBI studies, as could be done if the SKA operated as part of a VLBI array, it may be possible to detect splitting of the central image on mas-scales due to the combined effect of the black hole and central stellar cusp; in these cases all degeneracies can be broken and both components' masses can be measured separately.
\end{itemize}

\subsection{Studies of the lensed source population}  

Although the SKA will transform our understanding of the high redshift Universe, the sensitivity and angular resolution of the telescope can be increased by orders of magnitude (at no additional cost) through using the magnification provided by gravitational lenses. In this respect, almost any high-redshift object science application with the SKA could benefit from the use of natural telescopes, providing a model for the lens (the optics) is well understood; there are now several sophisticated methods available for the reconstruction of the distorted sources behind gravitational lenses (e.g. \citealt{wucknitz04b,vegetti09,hezaveh13}).

Here we briefly summarise some of the principle applications of the SKA strong gravitational lensing sample for studying the background source population.

\begin{figure}
\includegraphics[width=\textwidth]{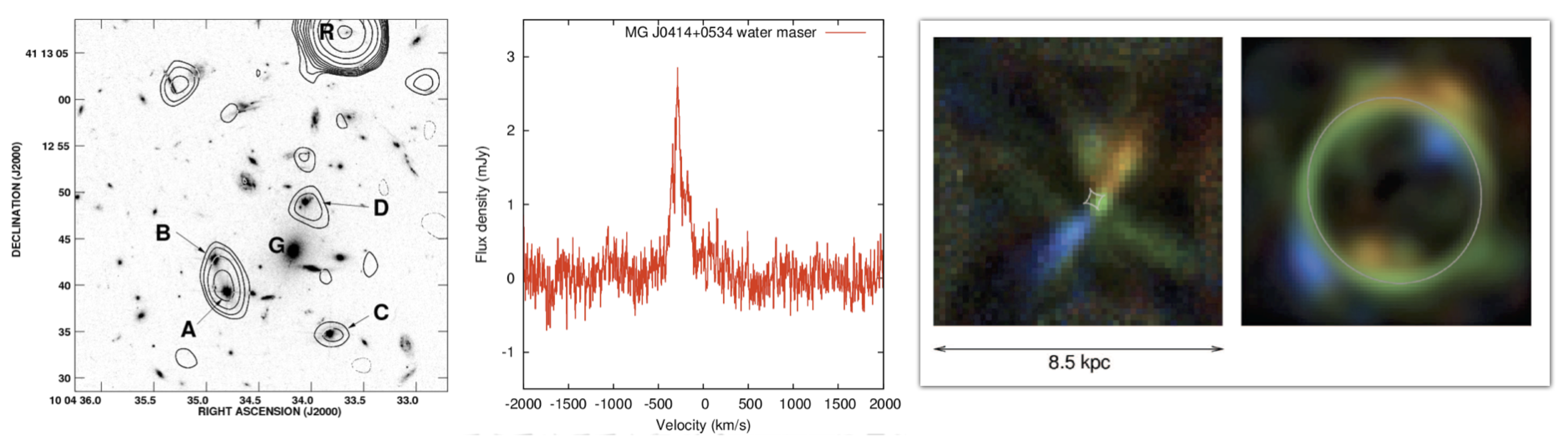}
\caption{(left) JVLA imaging of the highly magnified radio source (images A, B, C, D) behind the lensing cluster SDSS J1004+4112 \citep{jackson11}. (middle) The spectral line emission detected from the most distant known water maser galaxy, MG J0414+0534 at redshift 2.64 \citep{vio08}. (right) The CO (2--1) Einstein ring from the z = 4.12 quasar PSS J2322+1944 at 0.3 arcsec resolution with the VLA (right) and the reconstructed source (left) after lens modelling, which shows a rotation disk of gas \citep{riechers08}.}
\label{mag}
\end{figure}

\begin{itemize}
\item The increase in sensitivity (factors of 5--100) will allow a population of objects that could not otherwise be observed to be detected and studied in detail. For example, the faintest known continuum radio source ($S_{\rm 5~GHz} \sim 1~\mu$Jy; \citealt{jackson11}) is gravitationally lensed (see Figure \ref{mag}). The first immediate result will be to study the parent population of sources that will be detected with the SKA2 using the SKA1-MID array. The form of the parent population luminosity function is sensitive to the number of gravitational lenses that will be detected in the deep continuum surveys to be carried out with SKA1-MID. For example, about 440 gravitational lenses should be detectable in the SKA1-MID ultra deep survey (1 deg$^2$; rms 0.05~$\mu$Jy; see Prandoni \& Seymour 2014). Furthermore, the magnification provided by the gravitational lenses ($\sim 10$) will enable radio sources with star-formation rates of $>1$~M$_{\odot}$~yr$^{-1}$ at $z \sim 3$--4 and $>5$~M$_{\odot}$~yr$^{-1}$ at $z\sim6$ to be detected. These sources would directly probe the radio luminosity function in a regime that is around an order magnitude lower in luminosity than would be possible without the magnification from gravitational lensing.

\item The magnification will be particularly important for studying spectral line sources for which the large observational bandwidths will not increase the detectability, but will increase the volume that can be searched. An example is presented in Figure \ref{mag}, where the most distant radio spectral line in emission is shown (water maser at $z = 2.64$; \citealt{vio08}). Water maser systems are expected to be highly abundant at high redshift \citep{mckean11} and could be used to measure the masses of supermassive black holes in distant galaxies through observations of the maser line kinematics over time (e.g. see \citealt{kuo13} for studies at low redshift). Coupled with high angular resolution imaging (see below) of highly magnified water maser lines, it could also be possible to measure geometric distances for radio sources at moderate redshifts ($z\sim 0.7$--1) and test models for dark energy. In addition, water maser emission is known to be highly variable and so could also provide an independent measurement of dark energy through measuring the time-delay due to gravitational lensing (see below). High resolution observations would be possible using the SKA1-MID array Band 5 (probing $z\sim0.6$--4) for detection and variability monitoring, and with an SKA1-MID array that is part of a global VLBI array for imaging. Also, it may be possible to detect the HI absorption forrest from the Epoch of Reionization by targeting the population of high redshift objects that have had their intrinsic flux density boosted by gravitational lensing. Such observations would use the SKA1-LOW array.

\item The magnification provided by the lens increases the solid angle of the background source, which allows the object to be observed at high angular resolution. Such observations will be most useful for studying the morphology of radio sources at high redshifts and the structure of their various thermal and non-thermal components on pc to sub-kpc scales; e.g. star-formation, atomic and molecular gas distributions and AGN activity. An example of what can potentially be done is also shown in Figure \ref{mag}, where the CO molecular gas distribution of a redshift $\sim 4$ quasar \citep{riechers08} has been determined after correcting for the effect of the lens. This will require a high frequency capability ($<30$~GHz) to investigate high redshift galaxies ($z>3$), and the full collecting area of the SKA2 to study the formation of galaxies at the Epoch of Reionization ($z\sim6$--10).
\end{itemize}

\subsection{\bf Studies of cosmology}

Gravitational lenses can be used to measure distances at cosmological redshifts, and
are therefore potentially important for cosmology. The most direct application of gravitational
lensing for cosmology is the measurement of the Hubble constant, which can be
achieved by monitoring gravitational lenses with background sources that are variable in flux-density.
As the images result from multiple different light-paths, the measurement of
time delays between image variations gives a measurement of the difference in
angular diameter distance, and if the redshifts of the source and lens are known,
this in turn measures the Hubble constant, together with a secondary dependence
on other cosmological parameters such as $\Omega_m$ and $\Omega_{\Lambda}$. The first $H_0$ measurements were conducted on gravitational lenses discovered in the radio,
mainly using the VLA in the 1990s (e.g. \citealt{biggs99,fassnacht02}). Subsequent $H_0$ determinations were predominantly made using gravitational lens systems containing radio-quiet, optically-selected quasars, 
mainly due to the availability of dedicated small optical telescopes, for example, recent
measurements of time-delays have been dominated by the COSMOGRAIL collaboration
(e.g. \citealt{eulaers13}).

The major systematic problem with gravitational lens determinations of $H_0$ is
that, as well as the time delays, accurate mass models of the lensing galaxies
and the surrounding masses are needed. This process is subject to complex
degeneracies (e.g. \citealt{schneider13}), and needs
careful modelling and treatment of the systematics (e.g. \citealt{suyu10}) to achieve robust results. Nevertheless, two well-modelled gravitational lenses 
in which the systematics are under control give errors on $H_0$ (dominated by 
systematics) of about 5\% ($H_0 = 75^{+2.4}_{-3.6}$~km\,s$^{-1}$~Mpc$^{-1}$; \citealt{suyu13}), and further progress is likely in the next few years. 
Since $H_0$ is not determined by CMB measurements unless assumptions are made 
about other cosmological parameters (such as zero curvature or $w=-1$), very 
accurate standalone measurements of $H_0$ are capable of improving other 
cosmological constraints; for example, the dark energy figure of merit can be 
improved by factors of several \citep{linder11}, and the more reliable gravitational lensing 
results already provide a constraint of $w = -1.52^{+0.19}_{-0.20}$ at the 68 per cent confidence level \citep{suyu14}. 

The second major cosmological application of gravitational lensing uses the statistics of the numbers of
gravitational lenses or the properties of gravitational lenses in well-selected samples. These are sensitive both
to the numbers and masses of the potential lensing galaxies at particular redshifts 
(and hence to galaxy evolution), to the distribution and luminosity function
of the sources, and to the cosmological world model. Thus, we can use any 
samples in which the lenses and sources are well understood to constrain cosmology,
or use a world model, together with lens surveys to investigate galaxy evolution.
In the early days, gravitational lensing statistics were used to confirm the existence of
$\Lambda$-like terms in the world model (e.g. $\Omega_\Lambda = 0.69^{+0.14}_{-0.27}$; \citealt{chae02});
later, gravitational lensing statistics were used to provide evidence for (lack of) evolution
in elliptical galaxies \citep{chae03}. Recently \cite{oguri12} have
argued that lens statistics constrain $\Omega_{\Lambda}$ more robustly, $\Omega_\Lambda = 0.79^{+0.06}_{-0.07}~{\rm (stat.)}~^{+0.06}_{-0.06}~{\rm (syst.)}$, 
and are consistent with a $w=-1$ equation of state, although the latter requires the
combination with other cosmological probes.

The main cosmological applications of the SKA gravitational lens sample are as follows.

\begin{itemize}
\item Our simulations indicate that the SKA will provide a much more powerful probe of gravitational lenses in the high-redshift universe (see Figure \ref{sim}). On the most optimistic assumption, that of isothermal potentials in low-mass, high-redshift lensing galaxies, vast numbers of high-redshift lenses will be produced. The likelihood that this assumption is false, and that such galaxies will be more dark-matter dominated and hence less efficient gravitational lenses, itself implies that statistics of an SKA lensing survey will contain information about mass distributions in high-redshift galaxies. Improvements in the cosmological world model, in particular in $w$, are likely to result from the discovery and followup of a large number of new lens systems. In particular, the lensing statistics of samples from CLASS and the SDSS are based on samples of just 13 and 19 gravitational lenses, respectively. Increasing such samples by orders of magnitude, will result in the uncertainties being dominated by the systematics related to the assumed galaxy evolution model.

\item Increasing the samples of gravitational lenses by a factor of $>10^2$ will allow us to monitor gravitational lenses for time delays and be more selective in using those gravitational lenses for which the systematics can be well controlled. This will require the AGN-dominated systems to be observed and monitored with the SKA1-MID array in Band 5 to provide better data for gravitational lens modelling and to determine time-delay light curves (observations at 8.46 GHz at 0.2 arcsec resolution have been successfully applied in the past). Constraints on $w$ are then possible by combining the gravitational lensing time-delay data (which are mainly sensitive to $H_0$) and the CMB data from {\it Planck} (e.g. \citealt{suyu13}). Time-delay measurements will become more powerful both with numbers, and with a larger range of redshifts of gravitational lenses for which information can be extracted, as the statistical uncertainties are dominated by the handful of currently known systems that are appropriate for such an analysis.

\item Increasing samples by factors of $>10^2$ will also yield rare and cosmologically important objects. The ideal lens for cosmology would be one in which two sources are present at different redshift (e.g. \citealt{collett14}), but in which one is variable, such as a radio quasar. This allows an angular distance measurement without problems of mass degeneracy, providing kinematical information for the gravitational lens is known; a few such lenses give $w$ essentially free of systematic errors (see \citealt{schneider14} for discussion of the systematics). Such systems will be a factor 500-1000 rarer than ``standard'' radio lenses, because the lensing probability on any one line-of-sight is 0.1-0.2\%. This is within the likely yield of the SKA2 surveys.

\end{itemize}

\section{Summary}

The SKA era will transform our understanding of the Universe and strong gravitational lenses will play an important role in investigating galaxy formation and evolution over cosmic time, the properties of dark matter and the cosmological world model. This review has demonstrated that the SKA1-MID array is capable of detecting $\sim10^5$ new radio-loud gravitational lenses, providing there is sufficient sensitivity on 0.25--0.5 arcsec baselines and a frequency coverage between 1--14 GHz. A more conservative estimate suggests that  $\sim10^4$ should be straightforwardly identified in an SKA1-MID array survey at $\sim3~\mu$Jy~beam$^{-1}$ sensitivity and with a $3\pi$~sr sky area, with the main uncertainties being dominated by the luminosity function of the background source population. Coupling the SKA1-MID data with optical and infrared surveys, and further high resolution radio imaging, will be needed to identify the remaining gravitational lenses in the expected wide-area sky surveys. An initial pilot survey, matched to the sky area covered with {\it Euclid} should find several thousand new gravitational lenses from only 3 months of observing with the SKA1-MID array. Such a pilot survey would test the lensing statistics and lens identification methods. Finally, we note that it is important for several of the unique science goals of the SKA strong gravitational lensing programme that there is the VLBI capability of the SKA.

\end{document}